# Elemental Ferroelectric Topological Insulator in $\psi$-bismuthene

Yan Liang,[1,*] Xuening Han,[1] Thomas Frauenheim,[2,3,4] Fulu Zheng,[*,2] Pei Zhao[*,1]

[1]College of Physics and Optoelectronic Engineering, Faculty of Information Science and Engineering, Ocean University of China, Songling Road 238, Qingdao 266100, People's Republic of China;

[2]Bremen Center for Computational Materials Science, University of Bremen, 28359 Bremen, Germany;

[3]Beijing Computational Science Research Center, 100193 Beijing, People's Republic of China;

[4]Shenzhen JL Computational Science and Applied Research Institute, 518109 Shenzhen, People's Republic of China.

*Corresponding authors:

yliang.phy@ouc.edu.cn (Y.L.); fzheng@uni-bremen.de (F.Z.); zhaopei@ouc.edu.cn (P.Z.)

Ferroelectric quantum spin Hall insulator (FEQSHI) exhibits coexisting ferroelectricity and time-reversal symmetry protected edge states, holding fascinating prospects for inviting both scientific and application advances, especially in two dimensions. However, all of the previously demonstrated FEQSHIs consist two or more constituent elements. We herein propose the $\psi$-bismuthene, an uncharted allotrope of bilayer Bi (110), to be the first example of 2D elemental FEQSHI. It is demonstrated that $\psi$-bismuthene harbors measurable ferroelectric polarization and nontrivial band gap with moderate switching barrier, which are highly beneficial for the detection and observation of the ferroelectric topologically insulating states. In addition, all-angle auxetic behavior with giant negative Poisson's ratio and ferroelectric controllable persistent spin helix in $\psi$-bismuthene are also discussed. The emergent elemental FEQSHI represents a novel domain for both fundamental physics and technological innovation.

The discovery and exploration of novel materials have been at the forefront of scientific endeavors [1-5]. Among these materials, ferroelectric quantum spin Hall insulators (FEQSHI) have garnered immense attention [6,7], especially in two-dimensional (2D) lattices [8-12]. FEQSHI refer to a substance possessing a dual nature of ferroelectricity and band topological characteristics simultaneously, exhibiting both spontaneous electric polarization that can be switched by an external electric field [13] and robust gapless surface states that are topologically protected by time-reversal symmetry [14,15]. These multifaced attributes present profound implications for designing memory, computing, sensing and actuating with unparalleled control and efficiency [16].

Recent researches introduced the integrated ferroelectricity and band topology in 2D systems, ranging from monolayer [17-19] to multilayers [20-24]. Among which, either nonvolatile controllable nontrivial topological states or tunable edge helicity are demonstrated. Despite great progresses have been made, hitherto all these reported 2D FEQSHI involve multi-elemental compositions. This limitation primarily arises from the facts that, except for the scarcity of both 2D topological insulators [25] and ferroelectrics [26], it is conventionally believed that instigating ferroelectricity necessitates the multi-element interplay [27,28]. Fortunately, heavy Bi element, sits between metal and insulator in the periodic table, allows various allotropic transformations for the competing metallic and covalent bonding [29-31]. The rich competing structural phases and huge atomic spin-orbit coupling, providing possibilities to envisage 2D single-element FEQSHI. Indeed, a recent experiment discovered first elemental 2D ferroelectricity in $\alpha$-Bi [32], nonetheless, its band topology is still under passionate debate [33,34]. Exactly speaking, the explicitly study of single-element ferroelectric topological system has never been reported so far.

In this letter, we discover a new allotrope of two-dimensional bismuth, named as $\psi$-bismuthene, as the first intrinsic 2D elemental FEQSHI exhibiting simultaneously ferroelectricity and band topology. It showcases measurable nontrivial band gap and ferroelectric polarization, accompanied by a moderate ferroelectric transition barrier, which are beneficial for the experimental validations. Moreover, concurrent all-angle auxetic behavior with giant negative Poisson's ratio and ferroelectric controllable persistent spin helix in $\psi$-bismuthene are also demonstrated. These findings not only transcend the conventional boundaries of 2D FEQSHI, but also inspire further experimental investigations and technological developments.

First-principles calculations are performed with generalized gradient approximation Perdew-Burke-Ernzerh (PBE) functional [35], as implemented in the Vienna Ab Initio Simulation Package (VASP) [36]. The Brillouin zone is sampled with $7 \times 7 \times 1$ $k$ grids, and the thresholds of $10^{-5}$ eV and 0.01 eV/Å are adopted for energy and force convergences, respectively, at the plane wave cutoff energy of 500 eV. Phonon spectra are calculated with Phonopy code [37]. The ferroelectric polarization is examined by using Berry phase approach [38], and the corresponding barrier of ferroelectric reversing is evaluated via NEB method [39]. The topological features such as topological invariant, edge states and spin Hall conductivity are evaluated based on the maximally localized Wannier functions (MLWFs) that are constructed by Wannier90 interfaced with VASP [40].

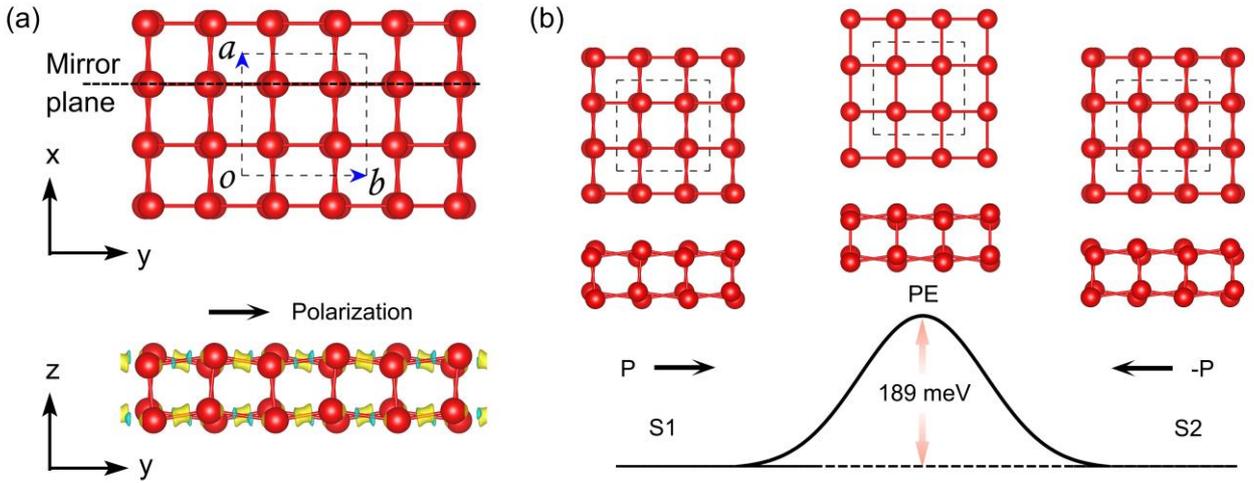

Figure 1. (a) Lattice structures of $\psi$-bismuthene from top and side views. (b) Energy landscape of structural switching process between energy-degenerate polarization states S1 and S2 across the transition state PE. Charge density difference between S1 and PE and the electric polarization is labeled in the side view of $\psi$-bismuthene.

Bismuth is the key element for constructing topological materials due to the huge spin-orbit coupling (SOC). Rooted predominately from the possibility to form hybridized and unhybridized bonds as well as the electron lone pairs, ultrathin bismuth film could adopt diverse allotropes such as $\alpha$, $\beta$, $\gamma$, and $\zeta$ forms [41,42], which further enriches intriguing prospects for exploring the interplay between competing structural phases and strong SOC. As shown in Fig. 1(a), herein we discover an additional allotropic form of 2D bismuth family, denoted as $\psi$-bismuthene. It can be deemed as a reconstructed bilayer of Bi (110) [29,43] with specific structural evolution that is distinct from previously reported Bi (110) bilayers in flat or distorted black phosphorus forms (DBP-Bi) [32,44]. Neighboring Bi atoms belong to different sublayers rearrange themselves, giving rise to nearest Bi-Bi interlayer bonding with identical bond length of 3.02 Å. Sublayer rearrangement features the resultant $\psi$-bismuthene an rectangular lattice with *ABM2* symmetry, in which the lattice constants are optimized to be *a* = 6.54 Å and *b* = 6.67 Å, respectively. To examine the structural stability of *ψ*-bismuthene, the phonon vibrations are calculated as depicted in Fig.S1. The absence of imaginary phonon modes across the entire Brillouin zone validates the structural dynamical stability of *ψ*-bismuthene. Additional calculations indicate that the binding energy of *ψ*-bismuthene is only 0.003 eV per atom higher than that of DBP-Bi. Such negligible energy difference indicates the comparable lattice steadiness between *ψ*-bismuthene and DBP-Bi, furtherly gives a strong hint for the experimentalists to detect the real existence of *ψ*-bismuthene.

In contrast to the mirror symmetric framework in *a* direction, it is easy to find that the Bi atoms in *ψ*-bismuthene is intrinsically off-centering displaced along *y* axis. Thus, noncoincident positive and negative charge centers and in turn the in-plane spontaneous electric polarization is intuitively expected. As presented in Fig.1 (a), this polarization is reflected by the isosurface of the difference in charge densities between *ψ*-bismuthene and a centrosymmetric structure PE [middle panel of Fig. 1(b)]. Clearly, charge accumulation and depletion regions deviate from each other, unambiguously confirming the appearance of in-plane electric polarization. Berry phase calculations reveals a net polarization in *ψ*-bismuthene reaches $1.06 \times 10^{-11}$ C/m, which is appreciable and comparable to other typical 2D ferroelectrics [45,46]. More interestingly, possible ferroelectricity is expected if taking the crystal symmetry into account. As schematically shown in Fig.1 (b), there are two energy-degenerate orientation states S1 and S2, which are closely linked to each other by an inversion operation. Obviously, in case they are switchable by an external electric field, *ψ*-bismuthene can certainly exhibit ferroelectric behavior with S1 and S2 correspond to two polarization states. To examine the feasibility of polarization states reversal, energy landscape of the structural switching process is analyzed by NEB method. It is found that the evolution between S1 and S2 goes through a centrosymmetric paraelectric state PE with a minimum energy barrier of 189 meV per unit cell. It should be noted that, the magnitudes of both the electric polarization and energy barrier are very close to the experimentally confirmed DBP-Bi [32,45]. These results firmly corroborate the preconception that *ψ*-bismuthene is a captivating 2D elemental ferroelectric candidate.

As the ferroelectricity and the mechanical properties of a crystal are mutually connected, hence the mechanical properties of *ψ*-bismuthene are also investigated. Two indicators, i.e., Young's modulus and Poisson's ratio are calculated and plotted in Fig. S2. It is found that both of these two indicators show anisotropic characteristics due to the structural anisotropy. The value of Young's modulus in the range between 34.57 and 155.58 N/m, reflecting the moderate rigidity against lattice deformation under external perturbations. Comparatively, it is rather interesting that *ψ*-bismuthene possesses all-angle negative Poisson's ratio (NPR). And remarkably, the giant NPR reaches an unprecedented value of -2.52, which is much larger than the currently available 2D NPR materials [47-50].

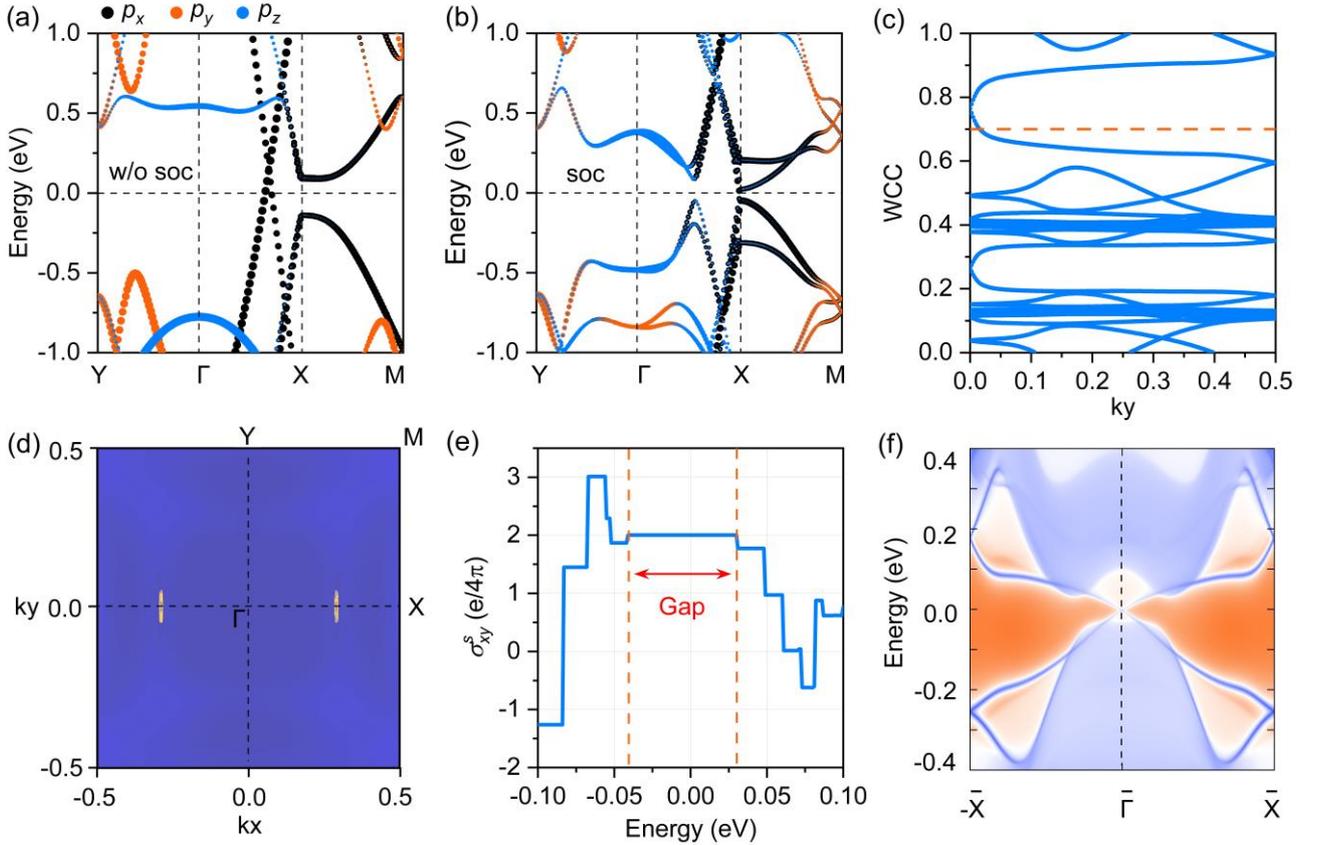

Figure 2. (a) Orbital-projected band structures of $\psi$-bismuthene (a) without and (b) with including SOC. (c) Evolution of WCC in the full closed plane of $k_z = 0$. (d) The $k$-space distribution of spin Berry curvature of the occupied bands in $\psi$-bismuthene. (e) The spin Hall conductance as a function of chemical potential, showing a quantized value within the SOC energy gap. (f) MLWF based topological edge states of the $\psi$-bismuthene along the [100] direction.

The electronic band structures of $\psi$-bismuthene based on PBE level are shown in Fig. 2(a-b). In the absence of SOC, it exhibits a metallic nature with a band crossing near the Fermi level along Γ-X direction, as can be seen in Fig. 2(a). In addition, it is indicated that such band crossing is mainly contributed by the in-plane $p_x$ orbitals. Such an origin of the crossed bands is totally different from other 2D bismuth films, among which the Dirac states are dominated by the out-of-plane $p_z$ orbitals [44, 29, 51-53]. When SOC is included, as displayed in Fig. 2(b), the energy degeneracy of the crossing point is lifted and $\psi$-bismuthene shows a well-defined insulator with indirect band gap of 72 meV. Furthermore, as a consequence of $M_y$ symmetry breaking, significant band splitting is noticed along Γ-X line.

Particularly, switching on SOC leads to the band gap opening, which normally indicates the nontrivial band topology. To judge the band topology of the ψ-bismuthene, we calculate the $Z_2$ invariant by analyzing the evolution of WCC. As shown in Fig. 2(c), the WCC is crossed by any arbitrary horizontal reference lines an odd number of times, directly giving a nontrivial $Z_2$ index ($Z_2$ = 1) for ψ-bismuthene. Beyond this evidence, the spin Hall conductivity ($\sigma_{xy}^S$) of ψ-bismuthene is obtained based on the Kubo formula [54]:

$$\sigma_{xy}^S = \frac{e\hbar}{2\pi^2} \int_{BZ} \Omega^S(\mathbf{k}) d^2k$$

The integrand $\Omega^S(\mathbf{k})$ is the spin Berry curvature of all occupied bands,

$$\Omega_n^S(\mathbf{k}) = -\sum_{n'\neq n} \frac{2Im\langle\psi_{n\mathbf{k}}|J_x^S|\psi_{n'\mathbf{k}}\rangle\langle\psi_{n'\mathbf{k}}|v_y|\psi_{n\mathbf{k}}\rangle}{(\omega_{n'} - \omega_n)^2}$$

Here $n$ is the band index, $J_x^S = 1/4(s_z v_x + v_x s_z)$ represents a spin current operator, $s_z$ denotes the spin operator, $\omega_n$ is the eigenvalue of Bloch wave function $|\psi_{n\mathbf{k}}\rangle$, $v_x$ and $v_y$ are the velocity operators. As displayed in Fig. 2(d), one can easily find that the spin Berry curvature $\Omega^S(\mathbf{k})$ distributes mainly in the vicinity of band crossing point and vanishes elsewhere. A quantized terrace of $\sigma_{xy}^S = 2e/(4\pi)$ within the energy window of the band gap is achieved by integrating the $\Omega^S(\mathbf{k})$ over the first Brillioun zone [see Fig. 2(e)], unrevealing the existence of quantum spin Hall (QSH) effect in ψ-bismuthene. As QSH effect is always accompanied by a hallmark of localized metallic helical edge states, we thus obtain the edge states of ψ-bismuthene based on the tight-binding Hamiltonian in the MLWF. Apparently, in Fig. 2(f), there is a clear edge Dirac cone dispersion connecting and conduction and valence bands with the Dirac point located at $\bar{\Gamma}$. Considering the possible underestimation of the band gap under PBE scheme, the nonlocal Heyd-Scuseria-Ernzerhof (HSE06) hybrid functional is further supplemented to double check the topological property, and it is found that the nontrivial band topology does not change. All these results manifest the ascription of ψ-bismuthene as a QSH insulator.

According to the discussions above, the simultaneous coexistence of ferroelectric and topological orders is therefore demonstrated in ψ-bismuthene. It is worth to note that, such an innate nontrivial feature is essential for protecting the ferroelectric order, since the microscopic depolarization could be prevented by the efficient charge dissipation at the metallic topological edges [55]. Most importantly, to the best of our knowledge, ψ-bismuthene proposed here represents the first example of 2D elemental FEQSHI with the confluence of ferroelectric and topological insulating features.

Different from centrosymmetric QSHIs, the nontrivial edge states at the opposite ferroelectric boundaries would be remarkably distinguishable due to the in-plane ferroelectric polarizaiton. To illustrate this point clearly, the edge states of $\psi$-bismuthene at [010] and [0$\bar{1}$0] boundaries are provided in Fig. S3. As expected, those two edge states are distinctly different. Absolutely, they will exchange upon the ferroelectric switching. This means the dispersion of the conducting edge at an assigned edge can be precisely manipulated by the ferroelectricity. Inversely, the detection features of edge states at the specific edge could serve as an alternative means to differentiate the polarization orientations.

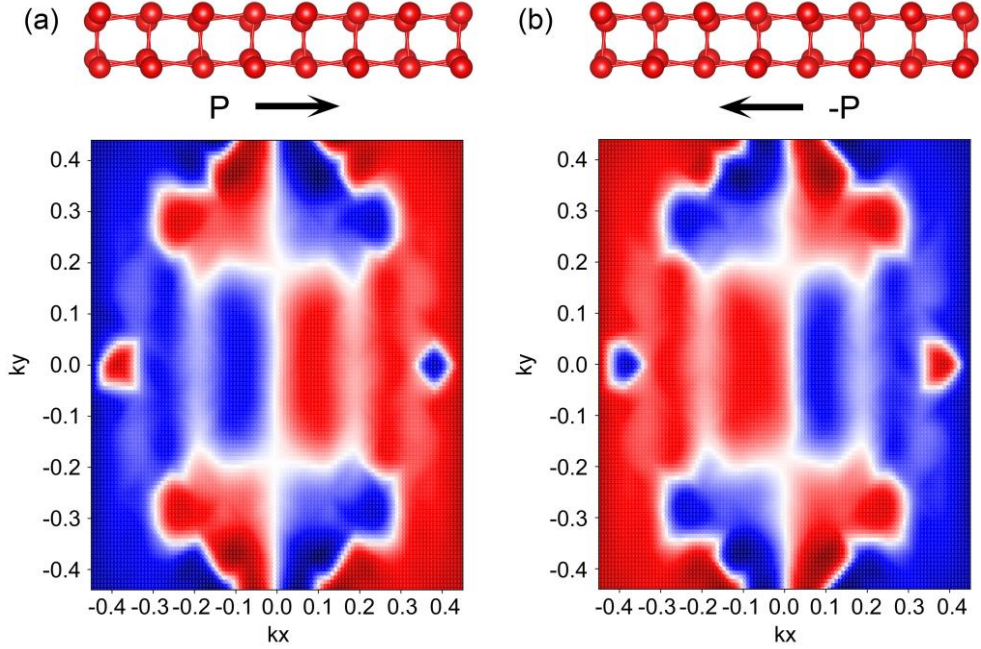

Figure 3. The spin polarization distributions for the highest valence bands of $\psi$-bismuthene within the first Brillouin zone under the opposite ferroelectric states (a) S1 and (b) S2, in which the in-plane spin textures are all zero.

The combined ferroelectric polarization and strong SOC commonly allows envisioning reconfigurable spintronic devices. For example, ferroelectric junctions and field effect transistors with added spin degrees of freedom. In this respect, we extend our study on the relationship between ferroelectricity and the spins of the nondegenerated bands in $\psi$-bismuthene. As a representative, the spin textures of the highest valence bands are plotted for $\psi$-bismuthene under S1 and S2 states. From the contour plots shown in Fig. 3(a-b), it is found that the spin polarization comes merely from the out-of-plane $S_z$, indicating the persistent spin helix in $\psi$-bismuthene. This originates from the *SU(2)* spin rotation symmetry of the systems belonging to $C_{2v}$ group symmetry [56]. Thus, symmetry protected unidirectional spin-orbit field is accessible in $\psi$-bismuthene, and the robust persistent spin helix (PSH) is immune to all types of impurity-induced spin independent scattering. In addition, the mutated spin texture at Γ-X is coincide with the spin Berry curvature distribution. More interestingly, ferroelectric switching leads to the reversion of the PSH, combined with the topological edge conducting, endowing ψ-bismuthene as a tantalizing platform to be utilized in controllable high-performance spintronics.

In summary, we present theoretical discovery of the first 2D monoelement FEQSHI in a new allotrope of two-dimensional bismuth ψ-bismuthene, exhibiting simultaneously ferroelectricity and band topology. It shows measurable nontrivial band gap and ferroelectric polarization with moderate ferroelectric transition barrier, which are beneficial for the experimental validations. Moreover, all-angle auxetic behavior with giant negative Poisson's ratio and ferroelectric controllable persistent spin helix in ψ-bismuthene are also disclosed. These findings not only promise a fertile ground for exploring emergent physical phenomena and propelling innovative technologies, but also contribute to the ongoing evolution of the field of condensed matter physics and materials science.

ACKNOWLEDGEMENTS

This work is supported by the Qingdao Postdoctoral Science Foundation (Nos. QDBSH20220202095 and QDBSH20230102037), Shandong Provincial Natural Science Foundation of China (No. ZR2023QA070), Qingdao Natural Science Foundation (No. 23-2-1-8-zyyd-jch) and the Young Talents Project at Ocean University of China.